\documentclass[prd,aps,twocolumn,showpacs,a4paper]{revtex4-1}

\usepackage{latexsym,graphics,epsfig}
\usepackage{amsmath,amsfonts,amssymb,amsthm}
\usepackage[
    colorlinks,%
    linkcolor=blue,citecolor=red,urlcolor=blue,
]{hyperref}

\def\barr{\begin{array}}
\def\earr{\end{array}}
\def\ben{\begin{equation}}
\def\een{\end{equation}}
\def\bena{\begin{eqnarray}}
\def\eena{\end{eqnarray}}

\def\const{\rm constant}

\newcommand{\Rs}{\mathfrak{R}}  
\newcommand{\OMs}{{\bf \Omega}} 
\newcommand{\OMp}{K} 

\newcommand{\GR}{general relativity}

\newcommand{\R}{{\mathbb R}}
\newcommand{\C}{{\mathbb C}}

\newcommand{\maps}{\colon}

\def\stackto #1 { \, {\stackrel{#1}{\longrightarrow}}\, }
\def\stackTo #1 { {\stackrel{#1}{\Longrightarrow}} }

\newcommand{\SU}{{\rm SU}}

\newcommand{\SO}{{\rm SO}}
\newcommand{\so}{\mathfrak{so}}
\newcommand{\SL}{{\rm SL}}
\newcommand{\Sl}{\mathfrak{sl}}

\newcommand{\g}{\mathfrak{g}}
\newcommand{\h}{\mathfrak{h}}
\newcommand{\p}{\mathfrak{p}}

\newcommand{\Ad}{{\rm Ad}}

\newcommand{\define}[1]{{\it #1}}

\newcommand{\we}{\wedge}
\renewcommand{\L}{\pounds} 
\newcommand{\Dom}{d^{\perp}_{\Omega}} 

\newcommand{\half}{\frac{1}{2}}

\newcommand{\hepth}[1]{\href{http://arxiv.org/abs/hep-th/#1}{arXiv:hep-th/#1}}

\newcommand{\grqc}[1]{\href{http://arxiv.org/abs/gr-qc/#1}{arXiv:gr-qc/#1}}

\newcommand{\arxiv}[1]{\href{http://arxiv.org/abs/#1/}{arXiv:#1}}

\newcommand{\webpage}[1]{{\color{blue}}\url{#1}{\color{blue}}}

\begin{document}

\title{\bf Spontaneously broken Lorentz symmetry for Hamiltonian gravity}
\author{Steffen Gielen}
\affiliation{Max Planck Institute for Gravitational Physics (Albert Einstein Institute), Am M\"uhlenberg 1, 14476 Golm, Germany}
\author{Derek K. Wise}
\email{derek.wise@gravity.fau.de}
\affiliation{Institute for Theoretical Physics III, Universit\"at Erlangen--N\"urnberg, Staudtstr.~7/B2, 91054 Erlangen, Germany}

\begin{abstract}
In Ashtekar's Hamiltonian formulation of general relativity, and in loop quantum gravity, Lorentz covariance is a subtle issue that has been strongly debated.  Maintaining manifest Lorentz covariance seems to require introducing either complex-valued fields, presenting a significant obstacle to quantization, or additional (usually second
class) constraints whose solution renders the resulting phase space variables harder to interpret in a spacetime picture.  After reviewing the sources of difficulty, we present a Lorentz covariant, real formulation in which second class constraints never arise.  Rather than a foliation of spacetime, we use a gauge field $y$, interpreted as a field of observers, to break the $\SO(3,1)$ symmetry down to a subgroup $\SO(3)_y$.  
This symmetry breaking plays a role analogous to that in MacDowell--Mansouri gravity, which is based on Cartan geometry, leading us to a picture of gravity as `Cartan geometrodynamics.'  
 We study both Lorentz gauge transformations and transformations of the observer field to show that the apparent breaking of $\SO(3,1)$ to $\SO(3)$ is not in conflict with Lorentz covariance.
\end{abstract}

\maketitle

\section{Introduction and motivation}

Lorentz symmetry is a slippery topic in Hamiltonian formulations of general relativity and quantum gravity, for a simple geometric reason.  The standard first step in Hamiltonian gravity is to pick a spacelike foliation, in order to define time evolution.  Such a foliation gives a hyperplane distribution in the tangent bundle of spacetime, specifying the `purely spatial' directions at each point.  However, if we then perform a Lorentz gauge transformation, the spatial hyperplanes rotate in such a way that the resulting distribution is in general {\em nonintegrable}---it need not be the tangent distribution of any foliation.  Since the property of being a spacelike foliation is preserved only under very carefully chosen local Lorentz transformations, it is little wonder that introducing a foliation tends to obscure the behavior of a theory under local Lorentz symmetry. 

In this paper,  we suggest an alternative approach: we reformulate Hamiltonian gravity {\em without} any spacelike foliation.  Instead, we introduce a {\em field of observers} in spacetime.  Each observer naturally has an associated spatial hyperplane, but these hyperplanes need not be tangent to any foliation.  Physically, one may imagine the observer field as a cloud of dust filling all of space; our aim is then to describe the dynamics of general relativity from the perspective of the cloud, regardless of whether its velocity distribution is integrable.  Our perspective is thus similar to approaches such as the dust model of Brown and Kucha\v{r} \cite{kuchar} or Einstein-\ae\/ther models \cite{einstaet}, though our observers serve as idealized test particles, and do not couple to gravity.  While our methods could be applied to generalize the ADM formulation \cite{adm}, our focus here is rather on the Ashtekar--Barbero approach \cite{ashtekar,barbero}, which is the starting point for canonical quantization in loop quantum gravity.  Lorentz covariance in this framework has been a topic of particular confusion and debate, which is why we direct our attention here. 

In fact, in the Ashtekar--Barbero approach, and in the large body of work on quantum gravity that has stemmed from it, there is an additional reason that Lorentz symmetry is somewhat elusive: besides the local splitting of spacetime into space and time, there is a subtly related `internal' or algebraic splitting.  From the Lagrangian perspective, general relativity involves an $\SO(3,1)$ connection describing the Lorentzian geometry of spacetime.  Going over to a Hamiltonian picture, part of this spacetime connection should be viewed as an $\SO(3)$ connection describing the Riemannian geometry of space.  At least, this is the idea.  In practice, getting from $\SO(3,1)$ down to $\SO(3)$  historically required either complexifying the connection or maintaining a real connection but explicitly breaking Lorentz covariance by partial gauge fixing. Since the Ashtekar--Barbero formulation is the version of Hamiltonian gravity we propose to generalize, let us review these issues a bit further.  

In the original Hamiltonian formulation of Ashtekar \cite{ashtekar}, general relativity 
in four dimensions is cast in a form similar to $\SU(2)$ Yang--Mills theory, exploiting the role of the Lorentz algebra $\so(3,1)\cong \Sl(2,\C)$ as the self dual part of $\so(4,\C)$.  However, Ashtekar's formulation is most directly a theory of {\em complex} general relativity.  In particular, the spatial connection lives in the Lie algebra of {\em complexified} $\SU(2)$ and its conjugate momentum is a triad that lives in $\C^3$ rather than $\R^3$.  Recovering {\em real} general relativity in the Ashtekar formulation means imposing `reality conditions' that are especially awkward to handle in the quantum theory. 

The alternative formulation given by Barbero \cite{barbero} is based on a real $\SU(2)$ connection and is thus more amenable to quantization, but is not manifestly Lorentz covariant.  Unlike in Ashtekar's version, the connection can no longer be interpreted as a spacetime connection \cite{samuel}. As shown by Holst \cite{holst}, Barbero's formulation can be derived from the action
\ben
S[\omega,e]=\frac{1}{8\pi G}\int\kappa_{abcd}\,e^a\wedge e^b\wedge R^{cd}[\omega]\,,
\label{graction}
\een
as a function of the coframe $e$ and $\SO(3,1)$ connection $\omega$ with curvature $R$.  Here the indices $a,b,c,\ldots$ label the standard basis of $\R^{3,1}$, and $\kappa_{abcd}$ is a non-degenerate symmetric bilinear form on $\so(3,1)$,
\ben
\kappa_{abcd}=\half\epsilon_{abcd}+\frac{1}{2\gamma}\left(\eta_{ac}\eta_{bd}-\eta_{ad}\eta_{bc}\right)\,,
\label{kappa}
\een
invariant under $\SO(3,1)$, where $\gamma$ is known as the Barbero--Immirzi parameter. Up to an overall scale, (\ref{kappa}) is the most general quadratic form on $\so(3,1)$ with these properties \cite{wisesigma}.

Holst's analysis used the `time gauge' condition $e^0_i=0$, where $i$ denotes a spatial coordinate index, and defined
\ben
A^{ab}:=\omega^{ab}+\frac{\gamma}{2}{\epsilon^{ab}}_{cd}\omega^{cd}\,,
\een
finding that, {\it because of the time gauge condition}, only the $\so(3)$ part of $A^{ab}$---identified with Barbero's connection---has nonvanishing conjugate momentum.

Time gauge breaks manifest Lorentz invariance.  The Hamiltonian analysis {\em can} be performed without assuming time gauge, but then one finds second-class constraints, due to the mismatch that 18 momenta conjugate to the components $A^{ab}_i$ are functions of just 12 components $E^a_i$ \cite{peldanreview}. Second-class constraints are difficult to handle in the quantum theory; one can solve them by introducing a Dirac bracket, for which the connection in general does not self-commute \cite{alexandr1}, although one can redefine variables choosing certain parameters so that a self-commuting connection appears \cite{alexliv,geil}. One can also directly parametrize the solution to the second-class constraints by new variables \cite{barros} in which the Hamiltonian constraint takes a rather complicated form. An interesting related formulation recently given in \cite{higherdim} seems free of second-class constraints, but features additional {\it simplicity constraints}. We take the view that while quantization may therefore be as difficult as in the absence of second-class constraints, the resulting variables are somewhat harder to interpret in terms of spacetime geometry.

The issues mentioned so far all arise from the classical theory.  But besides these, there have historically been additional confusions in the quantum gravity literature, especially with regard to the {\em internal} algebraic splitting that is supposed to relate the Hamiltonian and Lagrangian pictures.  On the Hamiltonian side, one has loop quantum gravity, based on the Barbero formulation with gauge group $\SU(2)$.  Quantum states in this theory are described by spin networks: closed graphs in space, with edges labeled by $\SU(2)$ representations:
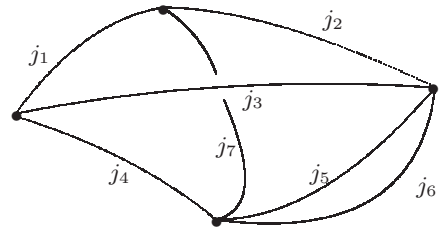
\begin{figure}[htp]
\centering
  \begin{picture}(200,100)
  \put(23,47){$\bullet$}
  \bezier{125}(25,50)(50,85)(80,90)\put(78,87){$\bullet$}
  \bezier{125}(80,90)(120,90)(182,60)\put(179,57){$\bullet$}
  \bezier{241}(25,50)(100,65)(180,60)
  \bezier{125}(25,49)(70,35)(100,10)\put(98,7){$\bullet$}
  \bezier{241}(100,10)(140,10)(182,60)\bezier{467}(100,10)(175,0)(182,60)
  \bezier{315}(80,90)(95,80)(100,65)\bezier{315}(103,55)(120,15)(100,10)
  \put(30,70){$j_1$}\put(140,83){$j_2$}\put(110,53){$j_3$}\put(60,25){$j_4$}\put(135,25){$j_5$}\put(175,20){$j_6$}\put(100,35){$j_7$}
  \end{picture}
\end{figure}

While these spin networks nicely describe the quantum geometry of space, viewing them as evolving in time prompted the introduction of spin foam models \cite{baez}.  Spin foams are state sum models proposed as the sum-over-histories counterpart to loop quantum gravity, and are described by 2-dimensional complexes with faces labeled by representations:
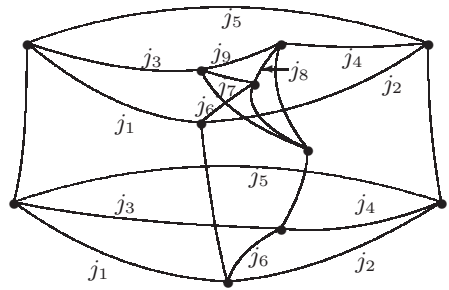
\begin{figure}[htp]
\centering
  \begin{picture}(200,110)
  \put(100,0){$\bullet$}\put(120,20){$\bullet$}\put(20,30){$\bullet$}\put(180,30){$\bullet$}
  \bezier{234}(102,3)(60,5)(22,33)\bezier{412}(102,3)(140,6)(182,33)\bezier{617}(102,3)(106,15)(122,23)
  \bezier{234}(22,33)(100,60)(182,33)\bezier{412}(122,23)(60,25)(22,33)\bezier{617}(182,33)(150,20)(122,23)
  \put(90,60){$\bullet$}\put(25,90){$\bullet$}\put(175,90){$\bullet$}
  \bezier{234}(92,63)(60,65)(27,93)\bezier{412}(92,63)(140,66)(177,93)\bezier{234}(27,93)(100,120)(177,93)
  \bezier{800}(22,33)(27,50)(27,93)\bezier{734}(182,33)(177,60)(177,93)\bezier{600}(102,3)(97,20)(92,63)
  \put(130,50){$\bullet$}\bezier{613}(132,53)(132,40)(122,23)
  \put(90,80){$\bullet$}\put(110,75){$\bullet$}\put(120,90){$\bullet$}
  \bezier{512}(132,53)(100,65)(92,83)\bezier{512}(132,53)(106,65)(112,78)\bezier{512}(132,53)(115,75)(122,93)
  \bezier{624}(112,78)(100,80)(92,83)\bezier{235}(122,93)(105,85)(92,83)\bezier{122}(112,78)(115,85)(122,93)
  \bezier{245}(112,78)(100,70)(92,63)\bezier{411}(27,93)(65,80)(92,83)\bezier{671}(122,93)(150,90)(177,93)
  \put(50,5){$j_1$}\put(60,60){$j_1$}\put(150,8){$j_2$}\put(160,75){$j_2$}
  \put(60,30){$j_3$}\put(70,85){$j_3$}\put(150,30){$j_4$}\put(145,85){$j_4$}
  \put(110,40){$j_5$}\put(100,100){$j_5$}\put(110,10){$j_6$}\put(90,67){$j_6$}\put(98,74){$j_7$}\put(125,80){$j_8$}\put(96,87){$j_9$}\put(125,83){\vector(-1,0){10}}
  \end{picture}
\end{figure}

The idea here is that a generic horizontal slice of such a spin foam should look like a spin network, and the labeled complex connecting two such slices is thought of as a higher-dimensional Feynman diagram with spin networks as initial and final states.  However, heuristic derivations of spin foam models start from the {\em Lagrangian} picture of classical general relativity, and it follows that the labels on spin foams come from the representation theory of $\SO(3,1)$, or rather its double cover $\SL(2,\C)$, {\em not} $\SU(2)$.  Evidently, slicing through a spin foam and getting a spin network involves {\em both} kinds of splitting we have been discussing: a geometric one that lowers the dimension of the complex, and an algebraic one that cuts down from $\SL(2,\C)$ to $\SU(2)$ representation theory.  For essentially this reason, the precise correspondence between the spin foam picture and the spin network picture was for a long time rather mysterious. 

Fortunately, it appears some headway has been made in recent years in the quantum theory, starting with the introduction of the EPR(L) and FK spin foam models \cite{eprlfk}. Like their predecessors, these models are based on the group $\SL(2,\C)$.  However, they also involve a choice of unit timelike vector in $\R^{3,1}$ for each edge in the spin foam, effectively selecting some $\SU(2)$ subgroup of $\SL(2,\C)$.  This leads to `projected spin networks' \cite{projected-spin-net} instead of the usual $\SU(2)$ spin networks.  $\SU(2)$ quantum states can be embedded into a Hilbert space based on $\SL(2,\C)$ in a way that keeps Lorentz covariance manifest, while at the same time clarifying the relationship to loop quantum gravity.  For a summary of this viewpoint see \cite{rovellispeziale} and references therein.  The observer fields discussed in the present paper may be thought of as the classical counterparts of the vectors attached to edges in spin foams or vertices in projected spin networks. 

Lorentz covariance continues to be investigated in high precision tests, e.g.\ using the gamma ray burst GRB090510 \cite{grburst} or neutrinos in the OPERA experiment \cite{opera}, and any serious theory of physics must prove itself consistent with such tests.  The consistency of a proposed quantum theory of gravity with these is ultimately to be decided at the quantum level by analyzing solutions to the dynamics.  While the EPRL/FK or other models may lead to a Hamiltonian quantum theory with appropriate Lorentz symmetry, it is hard to deny that one would feel safer starting from a classical theory where this symmetry is manifest. 

Our goal in this paper is to reformulate the canonical analysis of the action (\ref{graction}) in such a way that:
\begin{enumerate}
\item no foliation of space into spatial slices is needed, but only an arbitrary field of observers;
\item there is no need for second-class constraints or complexification, while at the same time Lorentz covariance is kept manifest;
\item there is a clearer geometric relationship between the {\em external} and {\em internal} splittings, providing an intuitive understanding of the apparent breaking of $\SO(3,1)$ to $\SO(3)$ at the classical, continuum level;
\item the Ashtekar--Barbero formulation is recovered as a special case, when the observer field comes from a foliation.
\end{enumerate}

The main new ingredient in our approach is the field of observers in spacetime. Given the coframe field, this can be turned into a field of `internal' observers: a field $y(x)$ taking values in the hyperbolic 3-space $H^3\cong\SO(3,1)/\SO(3)$ at each point in spacetime.  At each point $x$, $y(x)$ induces a splitting of $\so(3,1)$ into a subalgebra $\so(3)_y$ stabilizing $y$ and a complement $\p_y$.  The four-dimensional coframe field $e$ can be expressed in terms of $y$ and a triad $E$ which has only 9 independent components, and this allows for a fully covariant way to split the connection into spatial and temporal parts. Geometrically our constructions are best understood using Cartan geometry, describing the geometry of a spatial slice relative to a `model' $H^3$. We detail this construction in Sec. \ref{cartangeo}.
 
To our knowledge the results presented here have not been discussed before, but they might be subtly related to the framework of \cite{cianmont} which was also an attempt at a fully Lorentz covariant formulation of Ashtekar variables and loop quantum gravity. One of our motivations was to understand the results of \cite{cianmont} more clearly. For related work drawing connections between $\SU(2)$ loop quantum gravity and an $\SO(4,\C)$ covariant formalism see also \cite{alexliv}.

\section{Observers}
Our starting point in this paper is the action (\ref{graction}), so we have a coframe field $e\maps TM \to \R^{3,1}$ given from the outset, and we always assume it to be nondegenerate.   Using the standard basis of $\R^{3,1}$, the coframe gives us a basis of 1-forms $e^a$, orthonormal with respect to the induced spacetime metric $g_{\mu\nu}= \eta_{ab} e^a_\mu e^b_\nu$.

A \define{field of observers} is a unit future-timelike vector field $u$.  Using the coframe, we get the associated dual observer field, the unit timelike 1-form 
\ben
\hat{u}:= \,- e^a\,e_a(u)\,,
\label{defineuhat}
\een
where the Minkowski metric $\eta_{ab}$ is used to raise and lower $\R^{3,1}$ indices.  Physical fields given by differential forms split into purely temporal and purely spatial parts (denoted $\parallel$ and $\perp$), as seen by the observer, by
\ben
X^{\parallel}:=\hat{u} \we \iota_{u} X\,, \quad 
X^{\perp}:=X-\hat{u}\wedge \iota_{u} X\,,
\een
where $\iota_u$ denotes interior multiplication by $u$: it annihilates 0-forms, acts as $\iota_u X=X(u)$ on 1-forms, and is defined on higher forms by requiring it to be a graded derivation:
\ben
\iota_u(X\wedge Y)=(\iota_u X)\wedge Y+(-)^p X\wedge \iota_u Y\,,
\een
where $X$ is a $p$-form. In components, $(\iota_u X)_{\nu\ldots \rho}=u^\mu X_{\mu\nu\ldots \rho}$.
Since $\iota_u^2= 0$ and $\iota_u \hat{u} =1$ by construction, $\iota_u X^{\perp}=0$ for any differential form $X$.

We think of $\hat{u}$ as specifying a local `time direction,' and of the splitting of dynamical variables as generalizing the splitting done in the usual Hamiltonian formalism. 
We say the covector field $\hat{u}$ is \define{hypersurface orthogonal} if $\hat{u}=g\,df$ for some functions $f$ and $g$, or equivalently if $\hat{u}$ annihilates any vector tangent to a hypersurface $f=\const$.  By Frobenius' theorem, $\hat{u}$ is hypersurface orthogonal if and only if $\hat{u}\wedge d\hat{u}=0$. In the usual Hamiltonian formalism, $f$ is a time function, $\hat{u}=N\,dt$ where $N=1/\sqrt{-g^{tt}}$ is the lapse, and $u=(1/N)\left(\partial/\partial t+(g^{it}/g^{tt})\partial/\partial x^i\right)$.  One can for convenience always choose $u$ so that this is the case, though we emphasize that this is not necessary.

\section{Generalized Hamiltonian analysis of general relativity}
From the dynamical variables $\omega^{ab}$, a connection valued in $\so(3,1)$, and $e^a$, we define the observer-dependent fields by projecting into spatial and temporal parts, as described in the previous section:
\ben
\begin{array}{ll}
\Xi^{ab}:= \omega^{ab}(u)\,,& \Omega^{ab}:=\omega^{ab}-\hat{u}\,\Xi^{ab}\,; \\
y^{a}:= e^{a}(u)\,,& E^{a}:= e^{a}-\hat{u}\,y^{a}\,.
\end{array}
\een
An immediate consequence we will use in the following is that $E^a$ satisfies both 
\ben
E^a(u)=0 \quad \text{and}\quad y_a\,E^a=0\,.
\label{ortprop}
\een
Therefore $E^a$ is a {\it purely spatial 1-form valued in the 3-dimensional  subspace orthogonal to $y^a\in \R^{3,1}$}.

In order to express the curvature of $\omega^{ab}$ in terms of observer-dependent fields, it is useful to split the exterior derivative as:
\ben
d = d^{\perp}+d^{\parallel} \,.
\een
We think of $d^{\perp}$ and $d^{\parallel}$ as `spatial' and `temporal' differentials defined by the observer field. They are defined on any differential form $X$ by
\ben
  d^{\perp}X = dX -  \hat{u}\wedge \L_u X\,, \qquad  d^{\parallel}X = \hat{u}\wedge \L_u X\,,
\een
where $\L_u = \iota_u\, d + d\, \iota_u$ is the Lie derivative. 

Both $d^{\perp}$ and $d^{\parallel}$ are graded derivations, just as $d$ is. They do not in general square to zero:
\ben
(d^{\perp})^2 X = - d^{\perp} \hat{u} \wedge \L_u X\,,\quad 
(d^{\parallel})^2 X = d^{\parallel} \hat{u} \wedge \L_u X\,,\quad
\een
though these clearly vanish on any form $X$ that is \define{static} from the observer's perspective (i.e.\ $\L_u X = 0$). 
In fact, we do have $(d^{\perp})^2=0$ whenever the Frobenius condition is satisfied. To see this, note that from $\hat{u}\wedge d\hat{u}=0$ it follows that
\ben
d^{\perp}\hat{u}=d\hat{u} - \hat{u}\wedge \L_u \hat{u}=\iota_u(\hat{u}\wedge d\hat{u})=0\,.
\een
Conversely, if $d^{\perp} \hat{u} = 0$ then $\hat{u} \wedge d\hat{u} = \hat{u} \wedge d^\perp\hat{u} = 0$, so the Frobenius condition can be rewritten simply as $d^{\perp}\hat{u}=0.$

With these definitions, the curvature of $\omega$ is
\bena
R^{ab}[\omega] & = & \Rs^{ab} +(d^{\perp}\hat{u})\Xi^{ab}\nonumber
\\&&+ \hat{u}\wedge\left(\L_u\Omega^{ab}+(\L_u\hat{u})\Xi^{ab}-\Dom\Xi^{ab}\right)\,,
\eena
where we have defined a `spatial curvature' $\Rs^{ab}:= d^{\perp}\Omega^{ab} + {\Omega^a}_c\wedge\Omega^{cb}$ and a `spatial covariant differential' $\Dom$ acting on an $\so(3,1)$-valued $p$-form $X$ by
\ben
\Dom X^{ab}:=d^{\perp}X^{ab}+\Omega^a{}_{c}\we X^{cb} - (-1)^pX^a{}_{c} \we \Omega^{cb}\,.
\een
The spatial and temporal  parts of $R^{ab}$ are apparent. Furthermore,
\ben
e^a\wedge e^b=E^a\wedge E^b + \hat{u}\wedge\left(y^a\, E^b-E^a\,y^{b}\right)\,,
\een
and one finds that
\bena
\kappa_{abcd}\,e^a\wedge e^b\wedge R^{cd}& = & d\left(\kappa_{abcd}\Xi^{cd}(\hat{u}\wedge E^a\wedge E^b)\right)\nonumber
\\&&+\kappa_{abcd}\hat{u}\wedge\left[E^a\wedge E^b\wedge \L_u\Omega^{cd}\right.\nonumber
\\ && + \Xi^{cd} \,\Dom\!\left(E^a\wedge E^b\right)
\label{new4form}
\\&& + \left.2 y^a\,E^b\wedge \left(\Rs^{cd} +d^{\perp}\hat{u}\,\Xi^{cd}\right)\right]\,.\nonumber
\eena

We can then rewrite the action (\ref{graction}) as
\bena
S & = & \frac{1}{8\pi G}\int \kappa_{abcd}\,\hat{u}\wedge\left[E^a\wedge E^b\wedge \L_u\Omega^{cd}\right.
\label{newact}
\\&&\left. + \Xi^{ab} \Dom\left(E^c\wedge E^d\right)+ 2y^aE^b\wedge\left(\Rs^{cd} +d^{\perp}\hat{u}\,\Xi^{cd}\right)\right]\nonumber
\eena
plus a boundary term which can be neglected if we are only interested in determining the local dynamics. In the usual canonical formalism, where $\hat{u}=N\,dt$, the first term specifies the symplectic structure and the other two terms give the Gauss, Hamiltonian, and diffeomorphism constraints of vacuum \GR, enforced by the Lagrange multipliers $\Xi^{ab}$ and $y^{a}$ \cite{holst}.

The action (\ref{newact}) defines a variational principle for general relativity in the following sense. The dynamical fields are $E^a, y^a, \Omega^{ab}$, and $\Xi^{ab}$, where $y^a$ is a function valued in hyperbolic space $H^3\subseteq\mathbb{R}^{3,1}$ and one imposes $y_a\,E^a=0$ everywhere. We view $\hat{u}$ as a fixed background structure and $u:=y^a\,{\bf e}_a$ where ${\bf e}_a$ is the frame field defined by $e^a({\bf e}_b)=\delta^a_b$ for $e^a:=E^a+\hat{u}\,y^a$. It then follows that 
\ben
E^a(u)=E^a({\bf e}_b)y^b=y^a(1-\hat{u}({\bf e}_b)y^b)=0
\een
since $\hat{u}=-y_a\, e^a$. Finally, one imposes the additional constraint that $\Omega^{ab}(u)=0$ to restrict the allowed configurations $\Omega^{ab}$.

The field equations resulting from variation of (\ref{newact}) with respect to the dynamical fields under those constraints must be the Einstein equations implying vanishing of torsion and the Ricci tensor since we have just redefined variables in (\ref{newact}).

It is worth mentioning that the spatial differentials $d^{\perp}$ appearing in (\ref{newact}) can be replaced by the usual differential $d$, as $\hat{u}\wedge d^{\perp}X = \hat{u}\wedge dX$ for any $X$. While $d$ is the natural operation on differential forms on spacetime, we view $d^{\perp}$ as more natural from the observer viewpoint emphasized here. Using $d^{\perp}$ also clarifies the relation to the usual Hamiltonian formalism, since e.g. $\mathcal{G}^{ab}:=\Dom\left(E^a\wedge E^b\right)$ is the analog of the usual Gauss constraint which only involves spatial derivatives (cf. Sec. \ref{consanal}).

\section{Internal observers}
The coframe field lets us easily switch between the observer field $u$ and $y^a$, a choice of unit timelike vector in $\R^{3,1}$ at each point in spacetime:  
\ben
u\mapsto y^a := e^a(u)
\,,\quad y^a\mapsto u := y^a\,{\bf e}_a\,.
\een
We think of $y^a$ as the `internal' version of the observer field, as it plays a similar role: just as $u$ splits differential forms into spatial and temporal parts, $y^a$ splits $\SO(3,1)$ representations in an analogous way.  If $\SO(3)_y$ is the stabilizer of $y \in \R^{3,1}$, representations of $\SO(3,1)$ decompose into direct sums of $\SO(3)_y$ representations.  

For the fundamental representation and the adjoint representation, we have
\ben
\begin{array}{ccccc}
  \R^{3,1} &=& \R^3_y &\hspace{-.6em}\oplus\hspace{-.6em}&\, \R^1_y\,, \\[.5em]
  \so(3,1) &=& \so(3)_y &\hspace{-.6em}\oplus\hspace{-.6em}&\, \p_y\,.
\end{array}
\een
Explicitly, if $Y^a$ and $Z^{ab}$ are fields living in $\mathbb{R}^{3,1}$ and $\mathfrak{so}(3,1)$, respectively, then
\bena
{\bf Y}^a & := & Y^a + y^a\,y_b\,Y^b\,,\nonumber
\\{\bf Z}^{ab} & := & Z^{ab} + \left(y^a\,y_c\,Z^{cb}-y^b\,y_c\,Z^{ca}\right)
\eena
are valued, respectively, in $\mathbb{R}^3_y$ and $\so(3)_y$. Note that $y_a {\bf Y}^a = y_a {\bf Z}^{ab} = 0$. In general, this `internal' splitting will not be related to the spacetime splitting.  One case where they are related is the frame field itself: the spatial coframe $E^a$ already lives in $\R^3_y$, thanks to (\ref{ortprop}).

In the general case, applying both spacetime and internal splittings will give four different components. For the connection, one has the two splittings,
\ben
\omega^{ab} = 
\left\{\begin{array}{ll}
\Omega^{ab} + \hat{u}\,\Xi^{ab}\,, & \text{\footnotesize(spacetime)} 
\\[.5em]
 {\bf w}^{ab} - \left(y^a\,y_c\,\omega^{cb}-y^b\,y_c\,\omega^{ca}\right)\,. & \text{\footnotesize(internal)} 
 \end{array}\right.
\een
The spacetime and internal projections commute, so we can find the part of $\Omega$ that is both spatial and $\so(3)_y$-valued in either of two ways:
\begin{align}
\OMs^{ab}=&
\left\{\begin{array}{ll}
\Omega^{ab}+\left(y^a\,y_c\Omega^{cb}-y^b\,y_c\Omega^{ca}\right) & \text{\footnotesize($\so(3)_y$ part of $\Omega$)}  \\[.5em]
{\bf w}^{ab}-\hat{u}\,{\bf w}^{ab}(u) & \text{\footnotesize(spatial part of $\bf w$)} 
\end{array}\right. \nonumber
\\[.5em]  = & \;\; \omega^{ab} + \left(y^a\,y_c\,\omega^{cb}-y^b\,y_c\,\omega^{ca}\right)\nonumber
\\  & \quad -  \hat{u}\,\Xi^{ab} - \left(y^a y_c\,\hat{u}\,\Xi^{cb}-y^b y_c\,\hat{u}\,\Xi^{ca}\right).
\end{align}
Then by construction $\OMs^{ab}(u)=0=y_a\OMs^{ab}$, so that one can think of $\OMs$ as a spatial $\SO(3)_y$ connection.

Similarly, the complement of $\OMs$,
\ben
\OMp^{ab}=\Omega^{ab}-\OMs^{ab}\,,
\een 
is a spatial $\p_y$-valued 1-form.  

\section{Symmetries}

We can now consider two kinds of transformations:
\begin{itemize}
\item \define{Observer transformations}:  Make a new choice of spacetime observers, $u\mapsto u'$, with corresponding change in internal observers $y=e(u)$.  The fields $\omega$ and $e$ are not affected. 
\item \define{Gauge transformations}:  Perform a Lorentz gauge transformation in the usual spacetime sense.  The fields $\omega$ and $e$ transform as usual.  The observer field $u$ does not change, but its internal description $y = e(u)$ changes because $e$ changes.  
\end{itemize}
The first of these arises because general relativity clearly does not depend on an arbitrarily chosen observer field.
Behavior under the second kind of transformation is what is usually meant by `Lorentz covariance' in the quantum
gravity literature. We discuss each type of transformation in turn.

A change in observers can be achieved by a local Lorentz transformation, both internally and externally.  This works because the invertible coframe $e\maps T_x M \to \R^{3,1}$ at each point $x$ turns $T_xM$ into a representation of $\SO(3,1)$.  In particular, if $y^a\mapsto (y')^a = {\Lambda_b}^a y^b$ represents a change in the internal observer field, then  $\Lambda\in \SO(3,1)$ acts on $u\in T_xM$ by $u\mapsto \lambda u$, where $\lambda =  e^{-1} \Lambda e$. 
This gives a corresponding change $\hat{u}\mapsto \hat{u}\lambda^{-1}$, so that $\hat{u}(u)$ is invariant.  All timelike vector fields $u'$ are related to $u$ by some such transformation.  While the fields $\omega$ and $e$ are not changed, their splittings into temporal and spatial pieces of course do transform:
\ben
\begin{array}{lll}
(E')^a & = & E^a + \hat{u}\,y^a - \left(\hat{u}\lambda^{-1}\right){\Lambda_b}^a\, y^b\,,
\\[.5em](\Xi')^{ab} & = & \Omega^{ab}(\lambda u)+\hat{u}(\lambda u)\,\Xi^{ab}\,,
\\[.5em](\Omega')^{ab} & = & \Omega^{ab} + [\hat{u}-(\hat{u}\lambda^{-1})\,\hat{u}(\lambda u)]\,\Xi^{ab}
\\[.5em]&&\rule{.5em}{0em}-(\hat{u}\lambda^{-1})\Omega^{ab}(\lambda u)\,.
\end{array}
\label{observerchange}
\een
The action (\ref{newact}) is invariant under such transformations since it can be written as the action functional (\ref{graction}) of the fields $\omega$ and $e$. In general, for a given theory written in terms of observer-dependent quantities, invariance under (\ref{observerchange}) is a nontrivial property which is the analog in our framework of showing independence of {\em foliation} in standard Hamiltonian approaches. The transformations here form a much wider class since, as noted in the introduction, general changes of observer do not take foliations to foliations. One example of a framework not expected to be covariant under the change in local observer is the gravity theory proposed by Ho\v{r}ava \cite{horava}.

We now turn to gauge transformations in the sense of $\SO(3,1)$ gauge theory. Under local Lorentz transformations, a connection transforms as $\omega^{ab}\mapsto{\Lambda_c}^a\,\omega^{cd}\,{\Lambda_d}^b+{\Lambda_c}^a\, d\Lambda^{cb}$, and so 
\ben
\Omega^{ab}\mapsto {\Lambda_c}^a\,\Omega^{cd}\,{\Lambda_d}^b+{\Lambda_c}^a \,d^{\perp}\Lambda^{cb}\,.
\een
This looks like the formula for an ordinary gauge transformation of a spatial connection, given the interpretation of $d^{\perp}$ as a spatial differential. The $\SO(3)_y$ connection $\OMs$ transforms as $\OMs^{ab}\mapsto (\OMs')^{ab}$, where
\bena
\label{OMs-transf}
(\OMs')^{ab}&=&{\Lambda_c}^a\,\OMs^{cd}\,{\Lambda_d}^b + {\Lambda_c}^a\left(\eta^{cd}+y^c y^d\right)(d^{\perp}\Lambda){_d}^{b}
\\& = & {\Lambda_c}^a\,\OMs^{cd}\,{\Lambda_d}^b + \left(\eta^{ac}+(y')^a (y')^c\right)\Lambda^{d}{}_c(d^{\perp}\Lambda){_d}^{b}\,,\nonumber
\eena
where $(y')^a={\Lambda_b}^a y^b$. Note that $\eta^{cd}+y^c y^d$ is the induced metric on $H^3$ embedded into Minkowski space $\R^{3,1}$, and a projector onto $\so(3)_y$, so that $\OMs'$ annihilates $y'$.

Similarly, we see that under a Lorentz transformation
\ben
\OMp^{ab}\mapsto(\OMp')^{ab}={\Lambda_c}^a\,{\OMp}^{cd}\,{\Lambda_d}^b - {\Lambda_c}^a\,y^c y^d\,(d^{\perp}\Lambda){_d}^{b}\,,
\label{boostp}
\een
so that $\OMp'$ is in the complement $\p_{y'}$ of $\so(3)_{y'}$ and everything is covariant under $\SO(3,1)$. Under $\SO(3)_y$ transformations, $\OMs$ transforms as a connection while $\OMp$ lives in the representation $\p_y$, which is isomorphic to the fundamental representation of $\SO(3)_y$.

We have obtained a generalized Hamiltonian formalism where the local choice of vector in $\SO(3,1)/\SO(3)$ can be changed freely, similar to the one derived in \cite{barros}, but where we do not view $y^a$ as phase space variables. In components, if $u=(1/N)(\partial/\partial t+(g^{it}/g^{tt})\partial/\partial x^i)$,
\ben
y^a=\sqrt{-g^{tt}}\left(e^a_t+(g^{ti}/g^{tt})e^a_i\right)=(1/N)\left(e^a_t -N^i\, e^a_i\right)
\een
where $N$ and $N^i$ are the usual \define{lapse} and \define{shift} of canonical \GR\ familiar from the ADM formalism \cite{adm}. Here we follow the conventional treatment of lapse and shift, and hence the components of $y$, as Lagrange multipliers. We note that \cite{barros} parametrizes the choice of gauge by a 3-dimensional vector $\chi^I=-e^{It}/e^{0t}$, presumably using Beltrami coordinates on $H^3$, whereas \cite{alexandr1} defines $e^0_i=:\chi_I e^I_i$. Clearly one could use any set of coordinates on $H^3$ but in general the action of $\SO(3,1)$ will take a more complicated form in such coordinates. (Compare with the discussion for $\SO(4,1)$ in MacDowell--Mansouri gravity \cite{dgr}.)

\section{Cartan geometrodynamics}
\label{cartangeo}

In the `internal' picture, the field of observers simply picks a point $y(x)$ in hyperbolic space $\SO(3,1)/\SO(3)$, at each spacetime point, thus splitting our fields into various pieces, as we have seen.  This strongly resembles MacDowell--Mansouri gravity \cite{macdmans}, especially in its generalization by Stelle and West \cite{stellewest}, where (for positive cosmological constant) the enlarged gauge group $\SO(4,1)$ is spontaneously broken to $\SO(3,1)$ by picking a point in de Sitter space $\SO(4,1)/\SO(3,1)$, at each spacetime point, thus splitting an $\SO(4,1)$ connection into a Lorentz connection and a coframe field, to recover the action (\ref{graction}). 

Geometrically, MacDowell--Mansouri gravity and its Stelle--West reformulation are best understood in terms of {\em Cartan geometry}.  Since we have explained this in detail elsewhere \cite{wisesigma,cartan, dgr}, we review here just enough to compare to the present situation.   In this section, we show how our formalism can be viewed as \define{Cartan geometrodynamics}: a system of evolving spatial Cartan geometries, transforming equivariantly under gauge and observer transformations. 

In Cartan geometry, the geometry of an $n$-dimensional manifold $M$ is described relative to an $n$-dimensional {\em homogeneous} manifold called the `model space.'    The geometry of $M$ is then described via `rolling' the model space along paths in  $M$ without slipping---a process that is more strongly path-dependent the more the local geometry of $M$ deviates from that of the homogeneous model.   More precisely, if the model space has isometry group $G$, this `rolling without slipping' is described via holonomy of the \define{Cartan connection} on $M$, a $\g$-valued 1-form mapping tangent vectors to elements of the Lie algebra $\g$ of `infinitesimal isometries' of the model space.  This can be integrated along a path in $M$ 
to give a path in the configuration space of ways to place the model space tangent to $M$.  This path describes rolling without slipping. 
 
Essential to this `rolling' interpretation, however, is that Cartan geometry is invariant under gauge transformations of the Cartan connection---{\em but only under those gauge transformations that live in the stabilizer of the point of tangency} between $M$ and the model space.  If $y$ is the point of tangency in the model and $H_y$ is its stabilizer, the algebra $\g$ is reducible as a representation of $H_y$. In all cases of interest here, $G/H_y$ is a symmetric space (see e.g. \cite{wisesigma}) and hence $\g$ splits into a direct sum
\ben
\label{reductive}
   \g = \h_y \oplus \p_y
\een  
as $H_y$ representations.  This can be viewed as splitting the infinitesimal isometries $\g$ into those that preserve $y$ and those that translate $y$.  But {\em translating} $y$ is strictly forbidden if we are to roll the model geometry {\em without slipping}.   The no-slipping requirement thus breaks $G$ gauge symmetry down to $H_y$.  
In the Stelle--West formulation with $\Lambda > 0$, the splitting (\ref{reductive}), induced dynamically by a de Sitter space-valued gauge field $y(x)$, is what splits the $\SO(4,1)$ connection into the $\SO(3,1)$ connection $\omega$ and coframe $e$. 

In the same way, in our Hamiltonian formulation, the hyperbolic space-valued field $y(x)$ gives us a splitting:
\ben
   \so(3,1) \cong \so(3)_y \oplus \p_y\,.
\label{algsplit}
\een
We have used this already to split the `spatial' connection as $\Omega^{ab} = \OMs^{ab} + \OMp^{ab}$, but this is {\em not} the Cartan connection we are interested in.  
Rather, we note that the `triad' $E^a$ can equivalently be viewed as a $\p_y$-valued 1-form $E^{ab}$, where 
\ben
   E^{ab} := y^a E^b - y^b E^a\,, \qquad E^b = - y_a E^{ab}. 
\een
One can check that $E^{ab}$ lives in $\p_y$, and that under a pure rotation $\Lambda\in \SO(3)_y$\,,
\ben
   \Lambda_c{}^a \Lambda_d{}^b E^{cd} = y^a(\Lambda_c{}^b E^c) - y^b(\Lambda_c{}^a E^c)\,,
\een
so that the correspondence $E^a \leftrightarrow E^{ab}$ gives an equivalence of $\SO(3)_y$ representations $\R^3_y$ and $\p_y$.

$\OMs$ and $E$ are natural ingredients for Cartan geometry modeled on {\em three}-dimensional hyperbolic space $\SO(3,1)/\SO(3)$.  However, even though they are purely spatial, meaning that $\iota_u\OMs$ and $\iota_u E$ both vanish, they do live on {\em four}-dimensional spacetime and, as we have seen, there need not be any extended notion of `space' in our observer-based framework.     Because of this, a precise Cartan-geometric understanding of the theory we have presented here requires a bit of care.

When $\hat{u}\we d\hat{u} = 0$, we know that $\ker \hat{u}$ can be integrated to a foliation, and in this case, $(\OMs,E)$ becomes a (reductive) Cartan connection on each spacelike slice.  
In cases where $\hat{u}\we d\hat{u} \neq 0$, while the spatial distribution is nonintegrable, we can still interpret $(\OMs,E)$ as giving a slight generalization of Cartan geometry.  Even without a foliation into spacelike hypersurfaces, one can always draw a {\em curve} tangent to the spatial distribution, starting out in any spatial direction.  Following such a totally spatial curve, the holonomy still describes rolling of hyperbolic space from one spatial hyperplane to another.   However, we must think of this as a {\em spatial} Cartan connection living on {\em spacetime}: since the notion of `space' itself is not integrable, attempting to come back to `the same' spatial point will generally give a timelike displacement.     

From the viewpoint of Cartan geometry, a {\em metric} geometry arises from the `rolling' motion itself, by declaring the rolling to be isometric.    The image to keep in mind is that of a ball rolling over a surface: the point of contact between the two traces out a path on each, and these paths clearly have the same length at any time.  In the present case, the length of a spatial path in spacetime can be measured via the corresponding path, or \define{development}, in hyperbolic space. This works because the spatial metric induced from $E$ is just the spatial metric restricted to the spatial distribution.  In particular, for any {\em spatial} vectors $v$ and $w$, i.e.\ $\hat{u}(v) = \hat{u}(w) = 0$, we have
\ben
\label{spatialmetric}
q(v,w):=
  \eta_{ab}E^a(v)E^b(w)= \eta_{ab}e^a(v)e^b(w) = g(v,w)\,.
\een

Finally, let us consider the symmetries discussed in the previous section.  A change of observers, $y^a\mapsto \Lambda_b{}^ay^b$ corresponds to changing the {\em field of basepoints} in Cartan geometry.  At each point, (\ref{reductive}) is a direct sum of $H_y$ representations, but it is also $G$-equivariant: 
\ben
\label{equivariant}
   \h_{gy} = \Ad(g)(\h_y)\, , \quad \p_{gy} = \Ad(g)(\p_y) \,,
\een
are the corresponding representations of the conjugate subgroup $H_{gy}=gHg^{-1}$, for any $g\in G$.  Such a change is an act of violence in ordinary Cartan geometry: it mixes up pieces in the `connection' and `coframe' parts of the Cartan connection and (in cases where the coframe induces a metric) deforms the metric geometry, possibly even causing it to become singular \cite{randono}. It will also generically map a torsion-free geometry to one with torsion, as observed in \cite{dgr}.  

In our case, however, the basepoint change $y\mapsto y'$ also gives a corresponding change $u\mapsto u'$ in the observer field and hence in the definition of space itself.  Thus, while components of the spatial Cartan connection $(\OMs, E)$ are mixed up, we are also changing our minds about which space the geometry is supposed to describe.  The
fields $\OMs$ and $E$ transform in a coherent way to describe, simultaneously for each choice of observer field $u$, the spatial geometry seen by $u$.  

Lorentz gauge transformations, the second kind of transformation discussed in the previous section, also mix up the parts of the Cartan connection according to (\ref{equivariant}), this time  {\em without} changing the observer field.  This would again seem like the sort of gauge transformation that is forbidden in a Cartan geometric interpretation.   In our case, however, the spatial coframe $E$ is derived from the spacetime coframe $e$, which also responds to a Lorentz gauge transformation.  In particular, it is easy to see that the spatial metric (\ref{spatialmetric}) is invariant under such transformations. 

\section{Constraint analysis}
\label{consanal}
To understand the dynamical structure of \GR\ in our formalism, we focus on the first term in (\ref{newact}) determining the symplectic form in the Hamiltonian theory,
\ben
S=\frac{1}{8\pi G}\int \kappa_{abcd}\,\hat{u}\wedge E^a\wedge E^b\wedge \L_u\Omega^{cd}+\ldots
\label{dynterm}
\een
Since $E^a\wedge E^b$ is valued in $\so(3)_y$, {\it only the components of $\Omega^{cd}$ in a 3-dimensional subalgebra of $\so(3,1)$ have nonvanishing conjugate momentum}. For $\gamma=\infty$, where $\kappa_{abcd}=1/2\epsilon_{abcd}$, the momentum conjugate to the $\so(3)_y$ part $\OMs$ is constrained to vanish, and only $\OMp$ is dynamical. 

In the general case, one can make the subalgebras (\ref{algsplit}) explicit by choosing local bases $J^{ab}_I$ ($I=1,2,3$) for $\so(3)_y$ and $B^{ab}_I$ for the complement $\p_y$, so that
\ben
\kappa_{abcd}J^{ab}_I J^{cd}_J=\frac{1}{\gamma}\delta_{IJ}\,,\quad\kappa_{abcd}J^{ab}_I B^{cd}_{J}=\delta_{IJ}\,,
\label{inprod}
\een
satisfying the algebra
\bena
&&[J_I,J_J]=-\epsilon_{IJK}J^K\,,\quad [J_I,B_J]=-\epsilon_{IJK}B^K\,,\nonumber
\\&&[B_I,B_J]=\epsilon_{IJK}J^K\,.
\eena
(By $\SO(3,1)$ invariance, (\ref{inprod}) may be verified for $y=(1,0,0,0)$.) Then the combination appearing in (\ref{dynterm}) as conjugate to $E^a\wedge E^b=:(E\wedge E)^I J^{ab}_I$ is
\ben
A^I := \OMs^{I}+\gamma\OMp^{I}\,,\quad \OMs^{ab}=:\OMs^{I}J^{ab}_I\,,\;\OMp^{ab}=:\OMp^{I}B^{ab}_I\,.
\label{barbero}
\een
$A^I$ is the Barbero connection taking values in a local 3-dimensional subalgebra of $\so(3,1)$ and transforming as a connection under $\SO(3)_y$ by the remarks below (\ref{boostp}). $\p_y$ transformations will not affect the components $A^I$, but merely change the components of $J^{ab}_I$ and $B^{ab}_I$, i.e.\ of the subalgebras $\so(3)_y$ and $\p_y$ embedded into $\so(3,1)$.

As in time gauge, (\ref{barbero}) does not transform as an $\SO(3,1)$ connection. This property is directly connected to the use of the Hamiltonian formalism. A local choice of time direction induces a spontaneous breaking of Lorentz symmetry down to a local $\SO(3)$ group; general relativity is not just a gauge theory but also includes the coframe field, a soldering form which translates between the fibers over the manifold acted on by Lorentz group and the tangent spaces to each point. We stress again that the issue of Lorentz covariance of the quantum theory can only be decided by analyzing the symmetries of a `ground state' solution. What we have shown here is that there is no conflict between the apparent necessity to break $\SO(3,1)$ down to $\SO(3)$ and Lorentz covariance; the breaking can be done in a fully covariant way using a gauge field encoding lapse and shift. Formulations involving second class constraints as in \cite{peldanreview, alexandr1} seem to add unnecessary complications to the Hamiltonian formalism; the coframe field can be expressed in terms of a non-dynamical gauge field $y$ and a triad $E^a$ with only 9 independent components. 

Completing the Hamiltonian analysis of (\ref{newact}), the apparent six constraints resulting from variation with respect to $\Xi^{ab}$, normally interpreted as Gauss constraints corresponding to local $\SO(3,1)$ symmetry, split into two sets. Their projection onto $\so(3)_y$ is
\ben
\hat{u}\wedge d^{\perp}_{\bf\Omega}(E^a\wedge E^b)\approx 0\,,
\label{cons1}
\een
(where only $\OMs$ appears); the component in $\p_y$ is
\ben
\hat{u}\wedge\left(\OMp^{ac}\wedge E_c\wedge E^b - \OMp^{bc}\wedge E_c\wedge E^a\right)\approx 0\,.
\label{cons2}
\een
(\ref{cons1}) determines ${\bf\Omega}$ to be the Levi-Civita connection of $E^a$, while (\ref{cons2}) is an algebraic constraint on $\OMp$. Substituting $\gamma\OMp^{I}=A^I-\OMs^I_{{\rm Levi-Civita}}[E]$ into (\ref{cons2}), one is left with three first-class constraints on $(A,E)$. This agrees completely with the derivation of Ashtekar variables in \cite{thiemann}, where $\OMp$ is identified with the extrinsic curvature.

Together with the constraints imposed by $y^a$ there are seven first-class constraints on 9 degrees of freedom, just as in the usual presentation in time gauge, which we did not find necessary to impose here.

\section{Outlook}

In deriving a set of variables for Hamiltonian \GR\ that transform covariantly under $\SO(3,1)$, we have introduced a classical formulation based on a local notion of `time direction,' interpreted as a local observer, and not necessarily related to any foliation of spacetime.  The result is very much in line with the formalism in current spin foam models, where an embedding of $\SO(3)$ into $\SO(3,1)$ is specified locally by a choice of unit normal. We feel this lends weight to the claim that loop quantum gravity is compatible with local Lorentz covariance. A similar construction was recently given \cite{aristdan} in the context of group field theory, including a unit normal vector as an argument in the quantum field that represents a vertex of a projected spin network.  The precise relationship of our classical theory to these spin foam and group field theory proposals deserves further study. Although we have focused on the case of four spacetime dimensions, our formalism does not essentially depend on the number of dimensions and should straightforwardly generalize to higher-dimensonal frameworks such as \cite{higherdim}.

One reason we find the observer-based formulation appealing is its flexibility.  For example, since observer fields exist in any time-oriented Lorentzian manifold, a formulation like the one presented here can be used to describe local time evolution even in the absence of global hyperbolicity, where no spacelike foliation is even possible.  We must admit that classical or quantum Hamiltonian dynamics for a general observer field without a foliation leads into uncharted territory, and may lead to difficulties not present in standard foliation-based formulations.   On the other hand, we emphasize that one may always perform an observer transformation, at least locally, such that the spatial distribution is integrable.  At the same time, the inclusion of nonintegrable cases makes behavior under Lorentz transformations manifest, which was our main purpose. 

These methods could also be applied to situations not covariant under the change in local observer, such as the gravity theory proposed by Ho\v{r}ava \cite{horava}.  In fact, while our observer field $u$ has served simply as a convenient way to describe time evolution of vacuum general relativity, in a more complete theory such a field may well play a physical role.  The observer field might conceivably be replaced by some dynamical matter field that couples in such a way as to select preferred local notions of space and time.  Several current approaches to understanding quantum gravity involve preferred spatial slicing, including not only Ho\v{r}ava gravity, but also causal dynamical triangulations \cite{cdt} and shape dynamics \cite{shape}.   Methods like those presented here may be a good way to understand how, from a spacetime perspective, the local anisotropy in such theories may arise dynamically.  Work on such ideas is in progress.

\subsection*{Acknowledgments}
SG thanks Daniele Oriti for helpful comments on an early version of the manuscript.  DW thanks Joshua Willis and Miguel Carri\'on \'Alvarez for helpful discussions. We also thank John Baez, Norbert Bodendorfer, Thomas Thiemann, and Andreas Thurn for comments.

\begin{flushright}
preprint AEI-2011-87
\end{flushright}

\end{document}